# Plasmonic crystals with highly ordered lattice geometries using continuous metal films on diatom bio-silica as both scaffolds and sources of tuneability

*William P. Wardley*[1]*, *Johannes W. Goessling*[1], *Martin Garcia-Lopez*[1] *

[1]International Iberian Nanotechnology Laboratory, Braga 4715-330, Portugal

E-mail: william.wardley@inl.int; martin.lopez@inl.int

**Abstract:**

Diatoms are microscopic algae found in all of Earth's water courses. They produce frustules, porous silica exoskeletons, grown by precipitation of silicic acid from water. Frustule components, known as girdles, from some diatom species also feature highly periodic pore arrays with properties on the same scale and quality of ordering as manufactured photonic crystals. Here, the girdles of two diatom species with various lattice constants are used as bio-silica scaffolds for plasmonic crystals, combined with metals chosen for spectral agreement between permittivity and photonic crystal properties. The use of frustules as scaffolds for plasmonic application has been tested earlier, as nano-porous features can enhance scattering and local hotspots, useful for sensing application or SERS. However, both the use of girdles and of continuous metal films are new approaches offering previously unseen plasmonic effects. Angularly resolved dispersions from reflection Fourier spectroscopy are provided for *Coscinodiscus granii* girdles, forming well-ordered square lattices, and *Coscinodiscus wailesii*, forming ordered hexagonal lattices, covered with a 40 nm film of silver or with 40 nm of gold, respectively. These were confirmed with optical modelling of the 3D systems, including both metal surface layers and internal frustule structures, allowing for calculation of dispersions and electric fields. Measurements are further compared with structures produced via Focused Ion Beam microscopy, demonstrating the advantages of girdles as photonic crystal scaffold and as an alternative to conventional plasmonic crystal fabrication. The results demonstrate that high-quality plasmonic crystal scaffolds from diatoms are available in nature, and their natural diversity can be applied in a controlled way to obtain plasmonic crystals with a wide range of behaviours and spectral responses.

## 1. Introduction

Plasmonic effects, the result of resonant interactions between incident photons and electrons in conductors, are frequently discussed as a means of enhancing optical effects due to their ability to confine light, enhance the local field strength and allow sub-diffraction effects [1]. Typically, these are sub-categorised as Local Surface Plasmons (LSPs), as seen in metallic nanoparticles[2,3] or holes in metal films[4,5], and Surface Plasmon Polaritons (SPPs), travelling modes found at the interface between a conductor and a dielectric[6,7]. These are often collectively referred to as Surface Plasmon Resonances (SPR). These effects allow a number of useful and exploitable properties related to field enhancement and hot electron effects, such as driving opto-chemical reactions[8,9] and facilitating photocatalysis [10,11]. In particular, the behaviour of nanohole systems is more complex than might be expected at first glance due to resonant coupling between plasmons at both metal interfaces plus possible SPR propagation within the nanohole itself. The extensive research performed into this area demonstrate phenomena such as extraordinary transmission [12,13], where the total amount of light transmitted by a metal screen perforated with subwavelength holes is significantly greater than the percentage area of the holes in the screen.

One structure of particular interest is the plasmonic crystal, an array of repeating elements in the surface of a metal film [14–16]. These structures can show a combination of phenomena, including both local plasmonic effects across the individual elements and resonant grating effects due to the Bragg crystal-like structure[17].

There are various scenarios where plasmonic gratings or resonant plasmonic surfaces are already seeing significant use. One of the more common application is in their use as substrates for Surface Enhanced Raman Spectroscopy (SERS), and increasingly resonant SERS (SERRS)[18–20]. Here, resonant plasmonic behaviour generates electric field hotspots at edges and apexes; this local field enhancement then significantly enhances Raman scattering, allowing the detection and analysis of weak signals due to, for instance, low volume or concentration of analyte, with detection levels in some cases down to a single molecule [21–23].

Fabrication techniques for plasmonic crystals typically rely on serial techniques, such as Electron Beam Lithography (EBL) or Focussed Ion Beam (FIB) microscopy [24–26]. These yield high-quality gratings relying on elaborate fabrication machinery and time-consuming protocols. Even large area techniques, such as nanoimprint, still typically rely on these nanofabrication techniques to generate precursor patterns [27,28]. For true large-area growth, self-assembly techniques can be implemented, but these also have disadvantages[29–31]. Bioinspired approaches to photonic devices have gained interest either as inspiration or directly as a scaffold, due to the cost-effective yet sustainable quality of nanostructure materials[32,33]. However, bioinspired plasmonic nanostructures have yet to see extensive use.

Diatoms are a diverse group of microalgae which form a porous silica exoskeleton (a frustule). They have been suggested as a source of materials for photonic or plasmonic applications due to the geometric nature of the porous structures frequently closely corresponding to the wavelengths of visible light[34–36]. While diatom frustules of different species form in a wide diversity of shapes, sizes and morphologies, they are typically formed of two structurally different silica components; valves and girdles. Valves form two hemispheres which are then accompanied by a number of pairs of girdles (Figure 1). The girdles (and valves) can feature complex nanostructures, both in terms of pores, but also of hollows, ridges and spines, as well as complex internal porous networks.

In certain species, those with which this paper is concerned (namely the species *Coscinodiscus granii* and *C. wailesii*), these girdle nanostructures feature highly ordered, very precise lattices of pores, reminiscent of photonic crystals. Both the lattice parameter (pore-to-pore distance) and pore diameter are highly maintained within one individual structure, showing high fidelity across the whole girdle, and the lattice parameter is highly conserved within the species. Given that the lattice parameter frequently corresponds to optical wavelengths, it has been hypothesised that the lattice performs a photonic role in metabolic functions of the cell, somehow aiding the microalgae to serve an optically derived advantage[37–39]. Experimentally, it has been shown that properties typically seen in laboratory-fabricated photonic crystals are also present in the frustules of certain diatom species, including other *Coscinodiscus* sp.[40].

Due to these highly ordered systems of surface nanostructuring, available over large areas (for instance, *C. granii* girdles have external surface areas of approximately 7500μm$^2$ per girdle covered with highly ordered square lattice of pores with a lattice parameter of 285 nm), it has been suggested that diatoms could be used as a natural source of photonic crystal material, where simple growth techniques can yield many more times the active area than current micro- and nanofabrication techniques allow. This paper takes this idea one step further, by adding a secondary processing step to apply metal to the girdle surface, thereby producing plasmonic crystals.

Previous work on combining diatom frustule parts and plasmonics has mostly focussed on local effects, either depositing nanoparticles onto the surface [41,42], electroless electroplating[43] or using self-assembly techniques to grow small metallic particles and islands on the surface of frustules[44]. Other groups have looked at using the complex internal structures of the frustules by depositing metal nanoparticles within the pore network and then using the resulting optical signal as a biosensor [45–47] or chemosensor[48] or as a SERS substrate[44,44,49,50]. All these systems present plasmonic properties by inclusion of metal nanoparticles into the biosilica matrix. Here it is demonstrated that metal coated girdles, with a continuous metal film instead of nanoparticles, can act as plasmonic-photonic crystals on which the SPP resonance is directly linked to girdle morphology. This procedure provides additional properties, combined with the ability to select spectral excitation ranges by careful diatom species choice based on their natural frustule diversity. Also, the use of continuous metal film has advantages over nanoparticle systems in a number of scenarios, such as when considering large area samples for multitarget or multiplexed sensors[51,52] or for integration into current SPR detector technology. Analysis is shown for optical measurements of girdles from two diatom species that, due to natural diversity, offer square and hexagonal lattices with different spatial parameters. These variations make such natural systems suitable to support SPRs in a wide spectral range and for diverse functions, particularly when combined with the selection of different metals for plasmonics in different spectral windows.

2. Methods

Diatoms were grown in artificial seawater medium under well-defined growth conditions, as described in [40]. Diatom cell propagation was checked using visible light microscopy (Figure 1a) and cleaned frustules were inspected with SEM for structural analysis, which confirmed that these *Coscinodiscus* sp. present girdles with high quality slab photonic crystal properties. Frustules produced by two different species were selected for their lattice parameters. *C. granii* (strain K1834, Norwegian Culture Collection of Algae) was selected for their square lattice with a lattice parameter of 285 nm and *C wailesii* (strain 1013/9, Culture Collection of Algae and Protozoa) was chosen for their hexagonal lattice with a lattice parameter of 330 nm. The sample preparation process is illustrated in Figure 1. Once cleaned, the organic material-free girdles were isolated then drop cast onto glass cover slips and dried overnight in a fume hood[53]. Once dry, the deposited material was pressed gently with a second cover slip. This causes the girdles to split into only 2 or 3 relatively large pieces and to lay on their thin edges with the pore network facing up (Figure 1c). Flattening is apparently possible due to particularly low Young's modulus based on its nano-porous composition, allowing for very elastic responses[54]. Inspection via SEM showed that the porous network has the same arrangement on both the internal and external faces and that the material lies flat on the substrate. Silver (Ag) and gold (Au) thin films with variable thickness were deposited using a Kenosistek multitarget confocal sputter system. Thicknesses were calibrated and roughness measured with Atomic Force Microscopy (AFM) measurement.

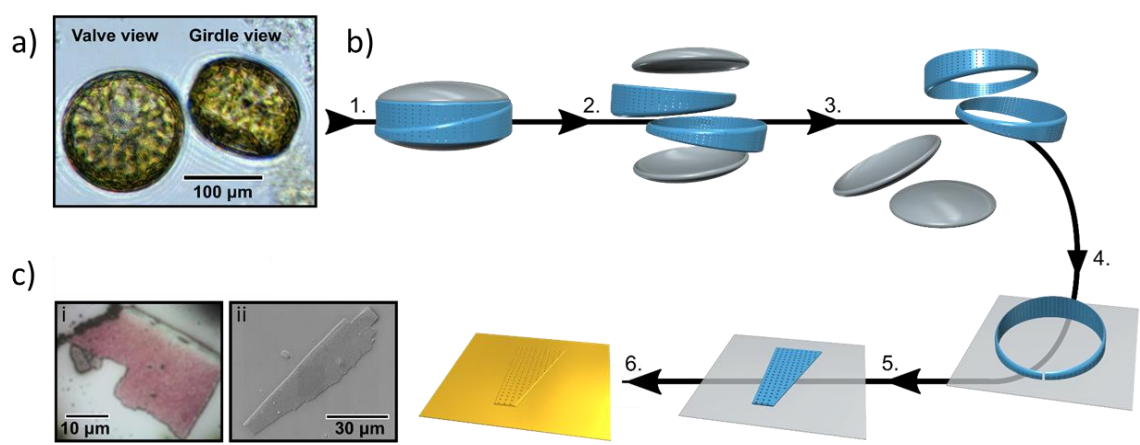

**Figure 1. Protocol for the preparation of high quality plasmonic crystals on diatom girdles.** a) Microscope image showing two living diatoms b) Flow-chart showing the sample preparation process. In live cells, girdles (blue) and valves (grey) are kept together by organic material. 2. After chemical removal of organic material, cleaned silica valves and girdles drift apart. 3. Separation of valves and girdles based on different microfluidic properties 4. Drop-casting girdle samples on glass cover slip and drying in air. 5. Breaking the girdles by application of pressure yield flattened pieces on glass substrates. 6. Bio-functionalization by deposition of defined thickness of noble metal. c) Microscope image (i) and SEM micrograph (ii) of metal-coated girdles from Coscinodiscus granii.

Nanohole plasmonic arrays are commonly inspected through reflectance measurements in which coupling to the plasmonic resonances is apparent as a strong reduction in the high reflectance of the metal films in the non-resonant spectral range[15]. Detailed optical characterisation of the metal-coated girdle was performed using an in-house built Fourier Image microscatterometry system[55], allowing for ≈10 μm spot diameter scanning and angle resolved spectroscopy without sample rotation. The backfocal plane image under white light epi-illumination was projected onto an imaging spectrometer (QImaging R6 Retiga CCD camera on a Princeton Instruments Acton Spectropro spectrometer) which allows recording the fully resolved k-vector/angle against wavelength by recording the whole image. The data is calculated as reflectivity relative to the reflection of a planar film of the same thickness as those deposited onto the girdles.

### 3. Results and discussion

*C. granii* was selected due to the high-quality lattice and photonic crystal properties of the girdle recently described for this species in detail [40]. Deposition of 40 nm Ag resulted in a closed thin film covering the surface, without entering the porous internal biosilica slab structure. As a result of the plasmonic effects studied in detail in this paper, it is easily observed by eye that the girdle band shows coloration in epi-illumination only on those areas with nanopores (Figure 1ci). To target high quality optical resonances in the structures, metal deposition was optimized based on numerical calculations for different metal thicknesses (see Figure S1). Quality and contrast of the spectral features were considered, in line with the lattice quality, which effectively decays when the metal layer closes over pores, thus removing the lattice. Based on this considerations, 40 nm Ag films sputtered onto the porous surface of flattened *C. granii* girdles were found to produce relatively sharp, distinct optical features without pore closing. This matched previously reported results for gold sputtering onto diatoms, where continuous films were deposited to generate SERS substrates, without features being lost or closed over[50].
As pore diameter can significantly vary between individuals in this species[53], a more
systematic experimental approach for identification of individual specimen morphology was
necessary to compare measurements and optical models. For this, indicative marks were
scratched into the surface of the metal film under the microscope in close proximity to selected single girdle specimen. Then, optical measurements were recorded prior to sample inspection using SEM (Figures 2a and 2b), where structural parameters of the whole array were extracted.

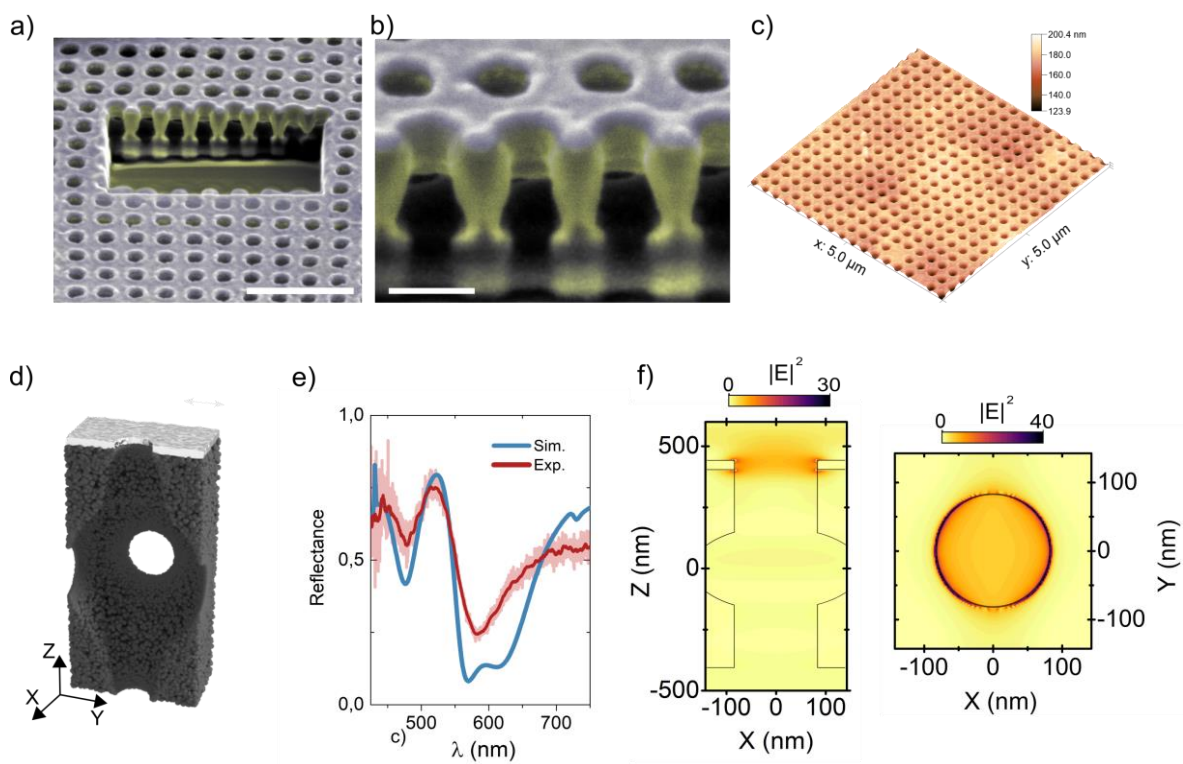

**Figure 2. SPP excitation at normal incidence.** a) Top view and b) detail of cross-section of *C. granii* girdle covered by 40nm Ag film. Scale bars 1 µm and 250 nm respectively. c) AFM of Ag metal surface showing metal film roughness after deposition. d) Sketch of half unit cell of the plasmonic crystal showing biosilica nano and mesoporous structuring (grey area) and thin metal layer (silver color). e) Comparison of experimental (red) and simulated (blue) spectra for normal incidence illumination. f) Electric field intensity distribution maps in cross section (XZ plan) and at metal surface (XY plan) extracted from calculations in (b) and for λ=560 nm. White lines shown as a guide to the eye for the geometry of the plasmonic lattice.

These measurements confirmed that the lattice parameter (centre-to-centre distance between the pores) in the girdle was $a = 285$ nm, arranged in square lattices with slightly different centre-to-centre distance in x and y, as previously described for this species[40].

In the case of one individual girdle specimen the pore diameter was $\Phi = 190$ nm. Note that previous studies have demonstrated that while pore diameter can vary strongly among specimens of the same species[53] the lattice parameter is highly preserved[40]. Surface maps derived from the AFM measurements highlight the quality of the plasmonic lattice as observed in the example in Figure 2c, confirming layers with the desired thickness and an RMS roughness of less than 5 nm for the surface of the metallic layer were deposited, ensuring the suitability of the metal films to support SPRs [47]. Numerical modelling of the whole structure (full girdle width and metal film) was performed to support the experimental results and aid in the attribution of the physical responses responsible for the optical effects observed. Due to the complex internal structure of the girdle (Figure 2b) an accurate model of the pore array was implemented with Finite Difference Time Domain numerical modelling (Lumerical FDTD Ltd) using morphological parameters extracted through high resolution electron microscopy images, with the resulting unit cell structure shown in Figure 2d. Experimental and simulated reflectance data at normal incidence (along z axis and polarisation along x) are shown in Figure 2e.

The two dips seen in the normal incidence reflectance spectrum ($\lambda \approx 480$ and 600 nm are due to the excitation of SPP resonances. For a perforated opaque metal film (thickness much higher than optical skin depth) the SPP excitation wavelength can be analytically calculated taking into account momentum matching for incident and SPP modes[5]. That model predicts excitation of the SPP at $\lambda_{spp} = 560$ nm for the determined lattice parameters of the grating at air-metal interface and normal incidence (see Figure S2 for analytically calculated plasmon dispersions of the analogous perfect lattice) representing a strong blue-shift compared to the experimental data, as well as to the rigorous simulations (FDTD) presented in Figure 2. In our case, the metal thickness (ω) is in the order of a few optical skin depths that, according to the well-established theory of plasmonic hole arrays[5], will enhance the coupling between the SPP modes at both air-metal and biosilica-metal interfaces. This will induce a redshift of the resonant wavelength. In this case the redshift is as large as ≈ 200 nm in accordance with previous studies[56,57]. Figure 2f shows the electric field maps for resonant wavelength λ = 560 nm calculated at the top of the metal layer and in cross section. The cross section (XZ plane) clearly shows the excitation of SPPs at both interfaces. Moreover, the perforated thin metal film will support short and long range SPPs[56] which are indicated by the dipolar character in the field distribution calculations at metal surface (XY plane) in Figure 2f.

In terms of the apparent additional dip in the reflection curve, seen at in the simulation but not in the experiment, this can be explained by examining the differences between the real and modelled systems. In the model, the metal film features sharp right-angles at both the silica and air interfaces, whereas the real system shows a much more rounded metal film growth, particularly at the air side. In addition, the modelling uses a perfect, infinitely periodic lattice with no deviation from an exact periodicity, whereas the girdle features natural irregularities and deviation from this perfect system. By introducing either curvature in the metal films or statistical deviation in the model, these two modes become less resolved, ultimately combining into a single mode (an example of this effect via numerical modelling for demonstrative purposes is shown in Figure S3). This means that in the real system a combination of rounded edges and small natural variation in pore diameter and position may produce a single reflection minimum rather than two cleanly resolved ones.

To address the full characterization of the plasmonic girdles, the optical analysis was extended to the full reflectance dispersions of the metal-coated girdles, with the equivalent calculated dispersions. Figure 3 shows the measured and simulated reflectance dispersions at incident angles $0 < \theta_{in} < 40$ degrees, along with associated electric field plots. The analysis was performed along ΓX direction of the square lattice and for both polarisations (pol) of the incident light (TM or p-pol (Figure 3a and 3b) and TE or s-pol (Figure 3c and 3d)). The agreement is outstanding, with a clear degeneration of the resonances at normal incidence for both polarisations at 2.21 eV ($\lambda$ = 560 nm). As larger angles of incidence are interrogated, highly dispersive resonances are observed for p-polarised illumination in accordance with the long range SPP excitation given that only p-polarised light will excite long range SPPs. However short-range propagating modes can also be excited by s-polarized beams as observed in Figure 3b. The non-dispersive response observed in reflectance for longer angles and s-pol is determined by the short-range propagating character of the excited plasmonic mode on this case. To corroborate this, field calculations were performed for resonances at $\theta = 21°$ and for two resonant wavelengths in each polarisation (Figures 3a and b, indicated by white annotation). As can be observed both polarisations show field concentration at air-metal and biosilica-metal interfaces but with a stronger field enhancement for the TM excitation as expected for a plasmonic system. The asymmetric profiles are a result of the excitation in a specific direction which, depending on the slope of the dispersion for particular wavelengths and angles, will correspond to a forward or backward travelling excitation (relative to the incoming k-vector of light) as observed for TM polarisation. In this configuration, $\lambda$ = 620 nm shows forward propagation as depicted by the negative slope of the dispersion which is also shown as strong field enhancement in the negative values of the X axis. Similarly, for $\lambda$ = 528 nm and same polarisation, the field enhancement is observed in the positive coordinates of the metal film corresponding to backward propagation as demonstrated by the positive slope of the dispersion measured and calculated in Figure 3a.

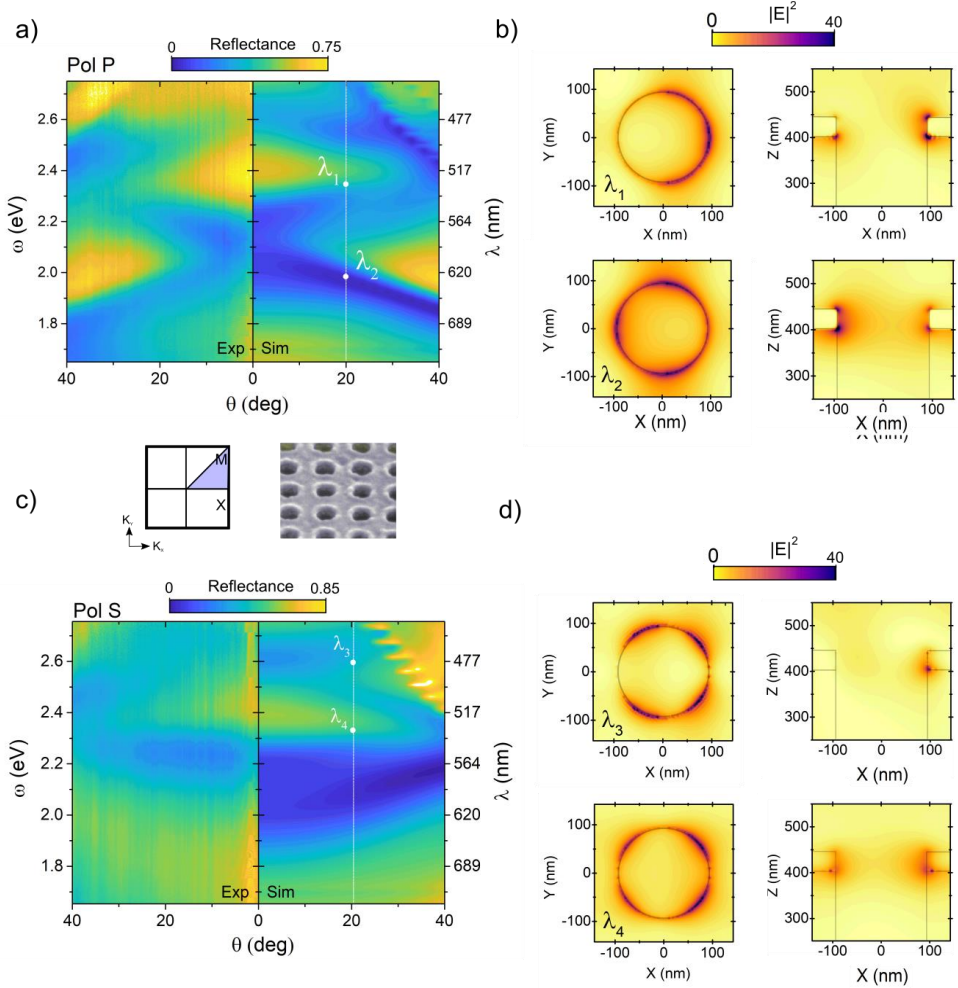

**Figure 3. SPP excitation along ΓX direction of the *Coscinodiscus granii* girdle plasmonic lattice.** a) Measured (left) and simulated (right) reflectance for the same metal coated girdle in Figure 2 under p-polarized illumination with incident angles 0°<θ<45°. b) Electric field maps for p-pol excitation with θ=21° and $\lambda_1$=620 nm and $\lambda_2$=528 nm as highlighted in (a). c) Reflectance under s-pol illumination. d) Electric field intensity maps for $\lambda_3$=534 nm and $\lambda_4$=488 nm and s-polarization. White lines delimit the unit cell geometry. Inset in (c) shows the Brillouin zone of the square lattice under study and three unit cells of a plasmonic girdle.

The quality of the plasmonic hole array based on biosilica scaffolds of diatoms was compared with a conventional fabrication technique, FIB nanopatterning using gallium ions to mill patterns that replicated those on the surface of the diatom girdles (FEI Helios NanoLab 450S DualBeam - FIB, 2kV, 3.2 nA beam current, producing a 51 x 51 hole array with a period of 285 nm). The total size of the structure was 14.25 µm², and thereby significantly larger than our interrogating spot (10.5 µm diameter). The resulting sample is presented in Figure 4a and its reflection dispersion in Figure 4b. Some important features of the SPP dispersion from the natural system could be repeated in the artificial fabrication; however, this system lacks the complex internal structure of the natural system with 3D porous systems, which could introduce differences in the plasmon formation and electrical field distribution.

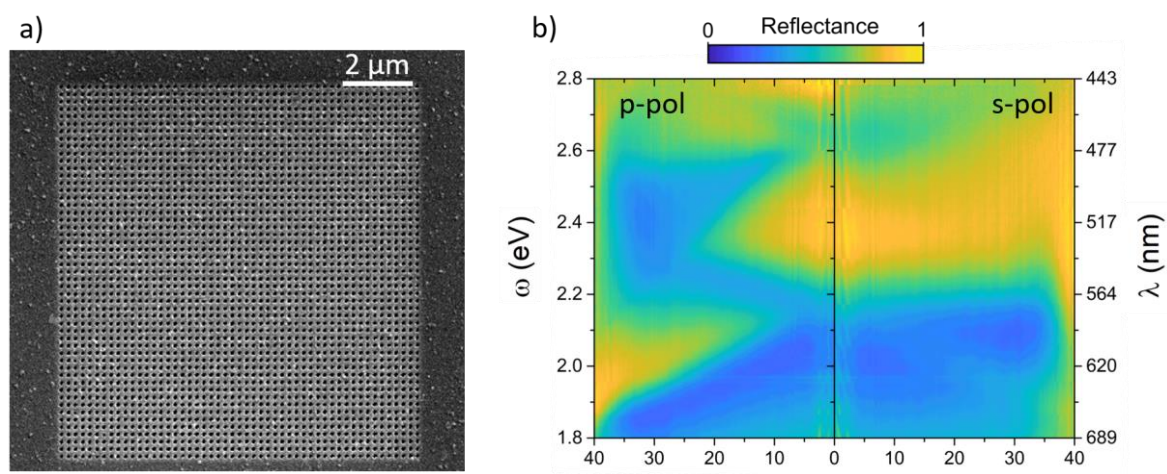

**Figure 4 Lattices fabricated with gallium ion lithography to compare to girdle-based plasmonic crystals**
a) SEM micrograph showing FIB fabricated lattice with the same lattice parameters as those found in *Coscinodiscus granii*. b) Angularly resolved reflectance dispersions for p- and s-polarised illumination of the lattice.

This experiment also demonstrates how disparate fabrication times and demands of both systems for plasmonic application are; the diatom scaffolds can be grown with minimal resources to very large volumes using basic marine growth media prior to chemical cleaning and metal deposition onto the surface, whereas the production of plasmonic-photonic crystals solely via nanofabrication techniques require access to high technology such as FIB microscopes. The latter was here done at patterning rates of approximate 100 holes per minute, with exact timings depending on material, thickness, etc. It is also worth reiterating that the diatom girdles feature a complex internal structure with pores in all three directions, in addition to meso-porosity of the biosilica slab, which cannot be mimicked with this artificial technique. These morphological differences in the dielectric scaffold account for both small changes in the observed optical response, but more importantly, add material properties to the natural system to be exploited or further explored (e.g. large internal volume and surface area among others).

While silver-coated girdles showed distinct plasmonic characteristics in the visible spectral range in the experiments presented above, silver itself is not always ideal for plasmonic application from a practical perspective; it is not suitable in the blue or ultraviolet due to intra-band transitions and it rapidly tarnishes due to chemical reactions with sulfur containing compounds in the atmosphere, converting it to silver sulfide (as seen in tarnished silver jewelry or cutlery), leading to loss of its plasmonic abilities[58]. By depositing alternatives, e.g. gold (Au), one can tune the plasmonic spectral characteristics, in this case to the red end of the visible or the near infrared light spectrum and increase the sample lifetime due to its chemical inertness and stability. However, pairing a photonic lattice scaffold that resonates with Au light absorption at around $\lambda = 550$ nm due to intra-band electronic transitions can limit the system at the blue spectral end. This is the case for the girdle of *C. granii*, with resonances around 500-600 nm in air. Both the distance between pores and the lattice types affect the optical resonance and consequently the electrical field distribution and SPP behavior of the plasmonic system. In particular, to more efficiently work at longer wavelength regions (the red and near infrared, $\geq$ 600 nm), a longer period and/or varied lattice type is required. However, as demonstrated before, these parameters are highly preserved in single diatom species (cite our paper here?) and cannot, so far, be controlled by biological growth control techniques during cell division. Therefore, as an alternative routine to achieve variation in the lattice constant of the diatom-grown plasmonic scaffold, a further diatom species, *C. wailesii*, was selected in this study. This species´ girdles forms hexagonal lattices (as opposed to square lattices in *C. granii*) along with longer period and larger pore sizes (330 nm and 160-200 nm diameter accordingly). The longer period causes red-shifted spectra for plasmonics, relative to the measurements in *C. granii*, meaning that the system is much better suited for the use of gold as the plasmonic metal. An SEM image of the surface showing the lattice is presented in Figure 5a, along with the measured and modelled dispersions for normal incidence (Figure 5b). Here the SPR excitation is evident as a dip in reflectance at $\lambda \approx 630$ nm, red shifted relative to the resonance seen in the *C.granii* square lattice (Figure 2).

To gain further insight into the excited SPR modes, angle and polarisation dependent reflectance along the crystallographic direction ΓK of the hexagonal lattice was measured (Figure 5c and 5d). Both the local plasmonic modes and the grating behaviour are clearly apparent in the p-polarised dispersions indicating short and long range plasmon excitation. For s-polarised excitation, the resonances obtained show reduced angular dispersion, indicating the predominant excitation of localized modes. Simulation results show correspondence with the measured results for both polarisations.

These results take the utility of this process even further. The desired spectral window can be targeted by selecting the diatom species with an inter-pore separation corresponding to the spectral region of interest, or by selecting a lattice type matched to the required function. Then, combining this selection with sensible material choice for deposition metals will yield high fidelity plasmonic modes at any desired UV/Vis/IR wavelength. The availability of hexagonal lattices with parameters suitable for VIS-NIR plasmonics is very interesting, as hexagonal lattices maximize filling fractions of pores, allowing for larger volume occupancy which could be interesting in applications such as biosensing. In terms of photonic properties, hexagonal lattices are more versatile in some photonic applications allowing for larger resonance shifts and polarisation dependencies between crystallographic directions than square lattices[57].

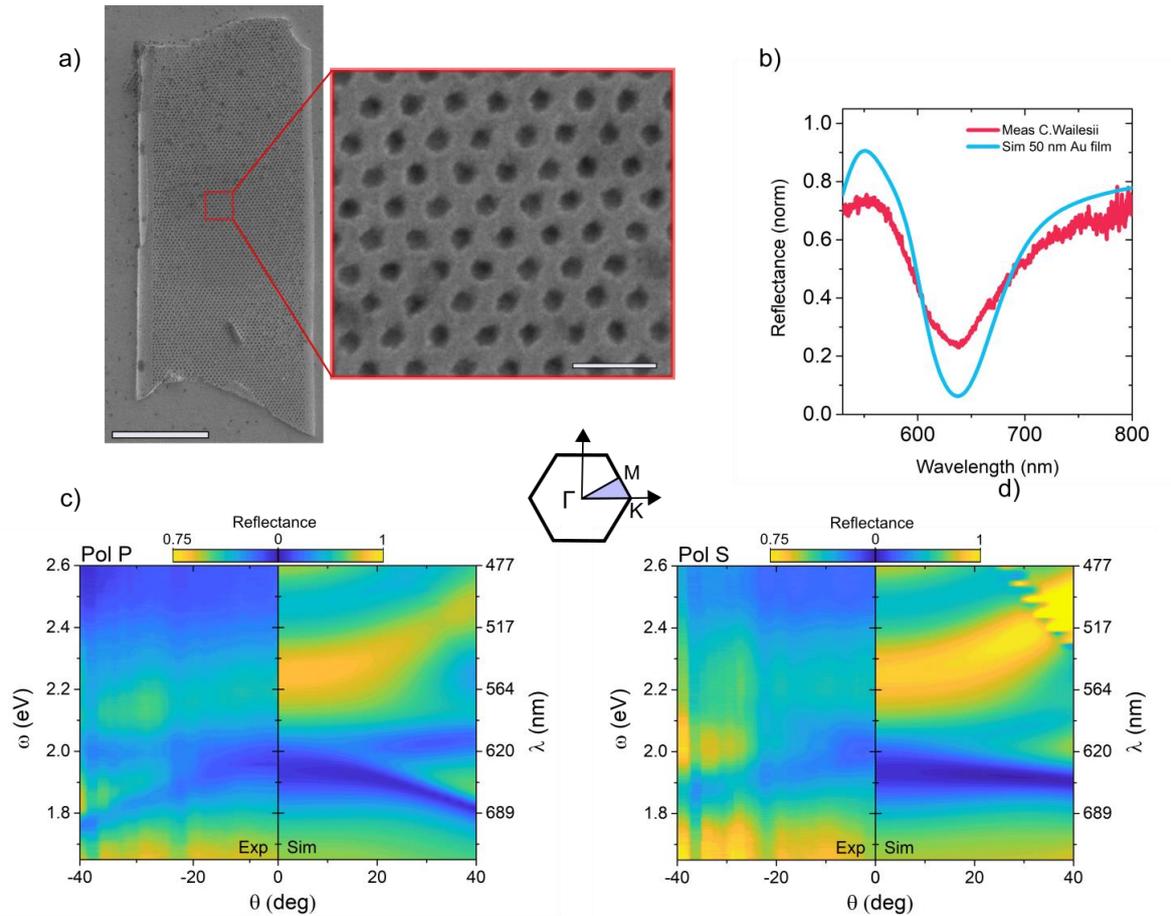

**Figure 5** *Coscinodiscus wailesii* **girdles as source of hexagonal lattice plasmonic crystals** a) Large area and detail SEM micrographs showing the lattice found in *Coscinodiscus wailesii*. Scale bars are 10 µm (left) and 1 µm (right) b) Normal incidence reflectance spectrum and simulated reflectance spectrum for illumination along ΓK. Reflectance dispersions for both experimental and simulated measurements for c) p-polarised and d) s-polarised illumination.

## 4. Conclusion

Diatom frustules have been used as scaffolds for plasmonic research in some earlier studies, but typically only via the addition of plasmonic nanoparticles, preventing the appearance of SPPs. The current study used diatom girdles with well-characterized lattice constants as plasmonic crystal scaffolds coated with continuous films of Ag and Au. The data presented shows that the girdles of diatoms are viable scaffolds for building plasmonic crystals with well-defined and reproducible spectral properties. Each scenario explored showed both the grating and local plasmonic behavior, while their specific modes were closely correlated with the theoretical counterparts calculated via numerical solvers, considering the complex internal structure of the biosilica material. Our results demonstrate that the most relevant photonic parameters in a slab photonic crystal, namely lattice type and lattice constant, can be selected by diatom species variety. In the special case of plasmonic crystals we have shown that by selecting diatom girdles with structural parameters that match the most optically suitable metal in terms of spectral response, it is possible to observe high fidelity plasmonic modes across the visible and near-infrared spectrum. The plasmonic behavior can be modulated by selecting a diatom structure featuring a porous network with particular lattice parameters and type which has been a long term missing component in naturally sourced photonic structures. Our results demonstrate that cultivated biosilica photonic nanostructures can offer a versatile and yet high-quality platform for applications in nano-photonics, of which plasmonics is only one.

Future possibilities for this system are to further explore the various parameters of the girdle and to expand the optical range into the UV. Additional tuneability of the optical response of the system can be found by controlling the diameter of the pores. While the pore diameter is well maintained within an individual, it is not conserved within a species, with different individuals showing significantly different diameters to one another. To control this source of natural variation, selective breeding and isolation techniques can be used to control these in larger numbers of individuals. In combination, these parameters allow for selection of specific properties in the visible and near-infrared spectral range. To extend this to higher energies, the UV would also be accessible by the use of diatom species with shorter girdle pore periods, in combination with sensible material choice - aluminium being the prime candidate, allowing a shift in the spectral response into the UV-spectral range due to its material properties[59]. The large variety of lattice structures available through diatom species diversity, with more than 100,000 estimated species[60], should be further explored as source of viable alternatives to expensive cleanroom fabrication that allows broad control of spectral properties of plasmonic crystals from an unconventional, globally available resource.

J.W.G acknowledges support and co-funding of the NanoTRAINforGrowth II program (project 2000032) by the European Commission through the Horizon 2020 Marie Sklodowska-Curie COFUND Programme (2015), and by the International Iberian Nanotechnology Laboratory. W.P.W and M.L.G would like to acknowledge the project POCI-01-0145-FEDER-031739 co-funded by Fundação para a Ciência e a Tecnologia and COMPETE2020.

# Supporting Information

**Plasmonic crystals with highly ordered lattice geometries using diatom bio-silica as both a scaffold and a source of tuneability**

*William P. Wardley\*, Johannes W. Goessling, Martin Garcia-Lopez\**

Figure S1      Metal thickness graphs

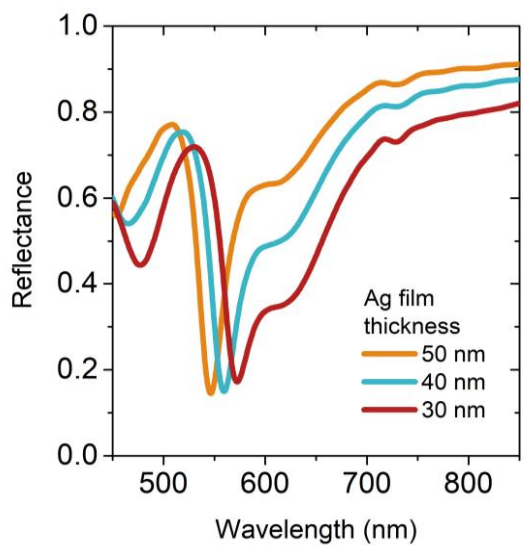

Graph showing the simulated normal incidence response for 3 different thicknesses of metal. This shows that the spectral response for 40 nm matches that of the *Coscinodiscus granii*. From experience of sputtering onto these samples, 50 nm resulted in significant closure of pores and 30 nm in practice resulted in films that were too thin to support plasmons efficiently based on empirical understanding of the system. Therefore, 40 nm was chosen as the working thickness.

Figure S2 Silver and gold plasmon dispersions

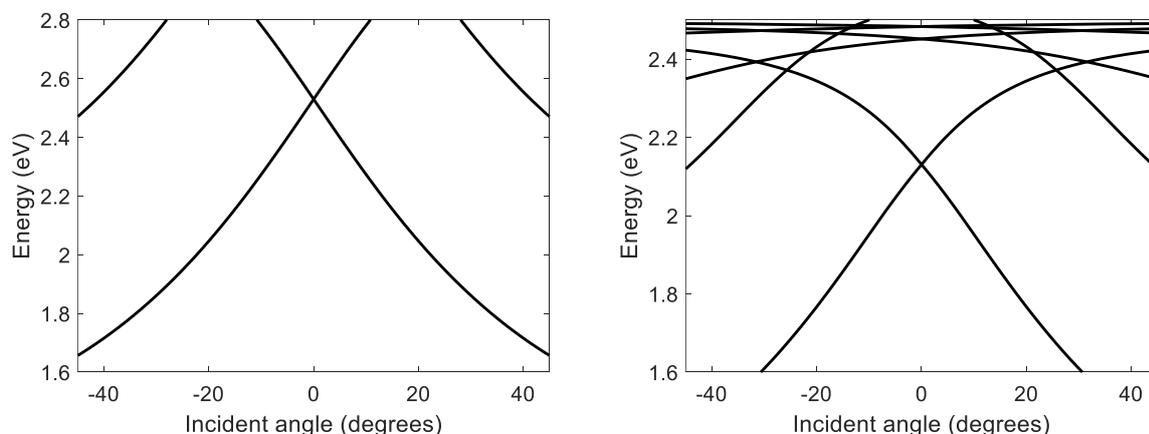

Theoretically calculated plasmon dispersions for a periodic perfect linear lattices of silver (McPeak et al., ACS Photonics 2, 326-333 (2015)) and gold (Johnson and Christie, Phys. Rev. B 6, 4370-4379 (1972)) in air and with periods 285 and 330 nm respectively. Note here we consider a semi-infnite corrugated metal-air interface. The differences with the resonant SPP excitation shown in Fig 3 and Fig 5 of the manuscript are due to the thin metal character of metal film over the diatom lattice which promotes coupling between SPPs at both air-metal and biosilica-metal interfaces.

Figure S3    Influence of radius of curvature on mode appearance

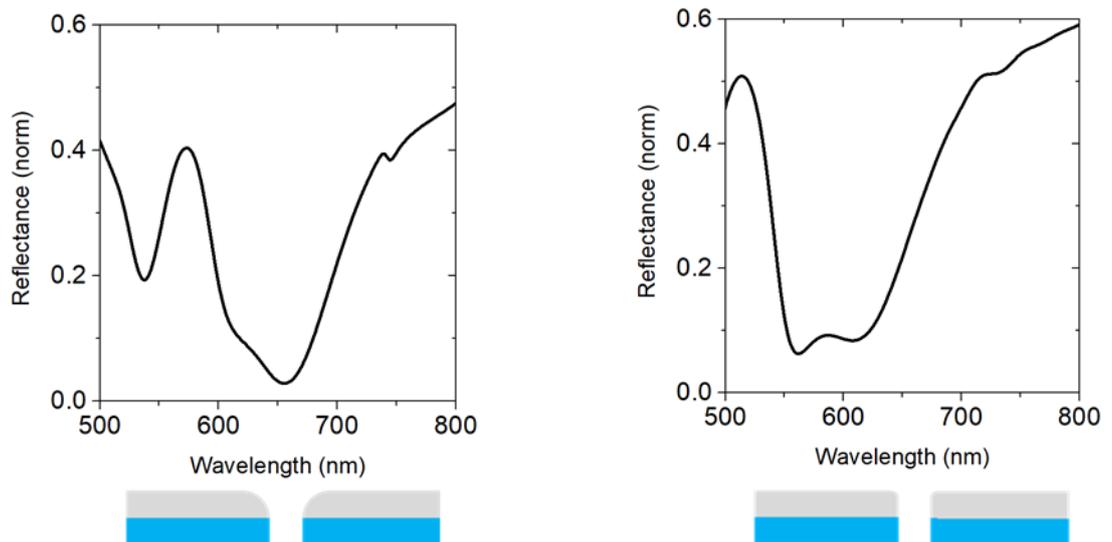

Influence of varying the curvature of the corners of the metal-air interface, illustratively showing the disappearance of two distinct reflectance minima into one single reflectance minimum. The wavelength shift is due to the effective increased radius of the upper side of the pore, which in the natural system will be seen as a broadening rather than a shift due to variations in both pore diameter and in radius of curvature.